\documentclass[twocolumn,showpacs,preprintnumbers,amsmath,amssymb]{revtex4}

\usepackage{graphicx}
\usepackage{graphics}
\usepackage{dcolumn}
\usepackage{bm}

\begin{document}


\title{Thermal transport properties of a charge density wave}
\author{Hiroyuki Yoshimoto }
\email{hiroyuki@kh.phys.waseda.ac.jp}
 
\author{Susumu Kurihara}%
\affiliation{Department of Physics, Waseda University 3-4-1 Okubo, Tokyo 169-8555, Japan}
\date{\today}
%
\begin{abstract}
Effects of collective modes on thermoelectric properties of a charge density system are studied. We derive the temperature dependence of thermoelectric power and thermal conductivity by applying the linear response theory to Fr\"ohlich Hamiltonian. Energy dissipation has been attributed to nonlinear interaction between phase mode and amplitude mode, ignoring disorder effects. We have found that the temperature dependence of the correlation function of electrical and heat currents is the same as that of electrical conductivity. This implies that thermoelectric power is inversely proportional to temperature. We have also found that temperature dependence of all the  correlation functions are essentially determined by the common mechanism -  nonlinear amplitude-phase interaction. Thermal conductivity has nearly constant value at the temperature above amplitude mode gap, and has exponentially low value at the temperature sufficiently below it.

\end{abstract}

\maketitle

\section{Introduction}
For many years, thermoelectric properties of various conductors have attracted both theoretical and experimental interest. These properties have also been investigated from practical standpoints. In this study, we are interested in thermoelectric transport properties of a charge density wave (CDW) system especially in connection with possible effects of collective modes.

As for electrical conductivity, a lot of studies on CDW system have been done until now. One of the dominant interest of such studies are focused on collective-mode contribution to conductivity. CDW has two collective modes- amplitude and phase modes. As is well known, phase mode can carry permanent currents in principle \cite{Frolich,Lee}. In practice, however, impurity and/or commensurability pinning prevent this phenomenon. Some effects of phase mode have been found in experiments of nonlinear conductivity and frequency dependence of ac-conductivity. Moreover, certain features of electrical dc-conductivity in CDW systems could be explained as many-body effects arising from nonlinear interaction between phase and amplitude modes  \cite{Kurihara1}. 

In contrast to electrical conductivity, theoretical study of thermal transport of this system has been rather rare. Experimentally, evidence of collective mode current of thermoelectric power have been found on nonlinear response to electric field \cite{Stokes,Kriza}. Some experimental data have also suggested fluctuation effect in thermal conductivity near transition temperature \cite{Kwok,Smontara,Kuo}. 
At present, it is not easy to observe collective-mode contribution in thermoelectric response, especially at low temperatures. This is because of pinning effect which is seen in electrical conductivity as well. However, in principle, these effect should exist in this region, and indirect evidences are expected to be actually observed. Theoretical study of this region is expected for better understanding of thermoelectric transports caused by collective modes.
For these reasons, we investigate the collective mode response to thermoelectric power and thermal conductivity of this system by applying linear response theory to a microscopic model. 
\vspace{5mm}
\section{Basic formalism }

We start from the Fr\"ohlich Hamiltonian
\begin{eqnarray}
H=\sum_{k}\epsilon_{k}c^{\dagger}_{k}c_{k}+\sum_{q}\omega_{q}b_{q}^{\dagger}b_{q} \nonumber \\ 
+\frac{\gamma}{\sqrt{N}}\sum_{q k}(b_{q}+b_{-q}^{\dagger})c^{\dagger}_{k+q}c_{k}.
\end{eqnarray} 
Here, electron-phonon interaction coefficient $\gamma$ are assumed to be constant equal to the values at $|q|=Q\equiv 2k_{F}$, since only the small portion of the phonon state are relevant in considering the CDW collective modes. We ingore electron spin indices for simplicity. Adopting Nambu representation\cite{Kurihara1,Kurihara2}, we can rewrite the above Hamiltonian as follows,
\begin{eqnarray}
H=\sum_{k}c^{\dagger}(k)(\xi_{k}\tau_{3}+\frac{\xi_{k}^2}{2\epsilon_F}\tau_{0})c(k)+\Omega\sum_{q}\sum_{j=1,2}b_{j}^{\dagger}(q)b_{j}(q)\nonumber \\
+\frac{\gamma}{\sqrt{2N}}\sum_{k,q}\sum_{j=1,2}[b_{j}(q)+b_{j}^{\dagger}(-q)]c^{\dagger}(k+q)\tau_{j}c(k),
\end{eqnarray}
where  $\tau_{1},\tau_{2},\tau_{3}$ are Pauli matrices,  $\tau_{0}$ is the unit matrix, $\epsilon_{F}$ is Fermi energy and
\begin{eqnarray}
c(k)&=&\left(\begin{array}{cc}
c_{k+Q/2} \\ c_{k-Q/2}
\end{array}
\right), \nonumber \\
\left(\begin{array}{cc}
b_{1}(q) \\ b_{2}(q)
\end{array}
\right)&=&
\left(\begin{array}{cc}
1/\sqrt{2} & 1/\sqrt{2} \\ -i/\sqrt{2} & i/\sqrt{2}
\end{array}
\right) \left(\begin{array}{cc}
 b_{q+Q} \\ b_{q-Q}
\end{array}
\right).
\end{eqnarray}
Summations are done in the range of $|k|<Q/2$ and $|q|<Q$ with respect to electron and phonon states.
The term $\frac{\xi_{k}^2}{2\epsilon_{\rm{F}}}$ is comes from the quadratic dispersion. Since transport phenomena are usually related to states near the Fermi surface,
linear dispersion relation is often adopted, espesially in one dimensional models. However, as far as thermoelectric transports are concerned, electron-hole asymmetry turns out to be important. It is because particles and holes respond to electric field in an opposite way: total energy flow vanishes if we have the exact electron-hole symmetry .

Electrical conductivity $\sigma$ is given as follows
\begin{eqnarray}
\sigma=\frac{e^2}{T}L^{11},
\end{eqnarray}
where $T$ is temperature and $L^{11}$ is transport coefficient calculated from current-current correlation function using Kubo-formula \cite{Kubo}.
Similarly, thermoelectric power S and thermal conductivity $\kappa$
are given by
\begin{eqnarray}
S=-\frac{1}{eT}\frac{L^{12}}{L^{11}},\\
 \kappa=\frac{1}{T^2}\large[L^{22}-\frac{(L^{12})^2}{L^{11}}\large],
\end{eqnarray}
where $L^{12}$ is calculated from correlation function between current and heat current operator, and $L^{22}$ correlation function between heat current operators, \cite{Langer}, \cite{Luttinger},and \cite{Kontani}, for example. 

Electric current operator is given by
\begin{eqnarray}
J=ev_{F}\sum_{k}c^{\dagger}(k)(\tau_{3}+\frac{\xi_{k}}{2\epsilon_{F}}\tau_{0})c(k).
\end{eqnarray} 
On the other hand, the heat current operator is composed of two parts: electron part and phonon part.
Historically, there are several forms that represent phonon part \cite{Mahan}. Such difference originates from various treatments of electron-phonon interaction. The following discussion is hardly affected by such differences. In this study, we give heat current operator of electron part as
\begin{eqnarray}
J_{Q}^{\rm{el}}= \hspace{70mm}\nonumber \\
v_{\rm{F}}\sum_{k,n}i(\epsilon_{n}+\omega_{l}/2)c^{\dagger}(k,\epsilon_n)(\tau_3+\frac{\xi_k}{2 \epsilon_{\rm{F}}}\tau_{0})c(k,\epsilon_n+\omega_l),
\end{eqnarray}
where, $\epsilon_{n}$ and $\omega_{n}$ are Matsubara frequencies for fermions and bosons. Note that, we need not to define heat current operator as $J_{Q}^{\rm{el}}-\mu J$ since we have already chosen the Fermi energy as the origen of energy. 
 Phonon part is also given by 
\begin{eqnarray}
&&J_{Q}^{\rm{ph}}=v_{\rm{ph}}\sum_{q>0,n}i(\epsilon_{n}+\omega_{l}/2) \nonumber \\&&\times[b_{q}^{\dagger}(\epsilon_n)b_{q}(\epsilon_n+\omega_l)-b_{-q}^{\dagger}(\epsilon_n)b_{-q}(\epsilon_n+\omega_l)],
\end{eqnarray}
which is rewritten in
\begin{eqnarray}
J_{Q}^{\rm{ph}}=v_{\rm{ph}}\sum_{q,n,i\neq j}i(\epsilon_{n}+\omega_{l}/2)b_{i}^{\dagger}(q,\epsilon_n)b_{j}(q,\epsilon_n+\omega_l) \label{phonon},
\end{eqnarray}
where, both $\epsilon_{n}$, and $\omega_{l}$ are boson Matsubara frequencies.
Using above operators, transport coefficient, $L^{ij}$ is given after usual analyitic continuation to real frequencies, by
\begin{eqnarray}
L^{ij}=\lim_{\omega\to 0}{\rm Im}{\Big(} \frac{\tilde{L}^{ij}(\omega+{\rm i}0_{_{+}})}{\omega}{\Big)},
\end{eqnarray}
where $\tilde{L}^{ij}({\rm i}\omega_n)$ is corresponding correlation funciton.

Here, We restrict ourselves to low temperature region $T<<T_{\rm{p}}$ in the following discussion.  
 Peierls gap $\Delta$ defined by
\begin{eqnarray}
\Delta=\frac{\gamma}{\sqrt{N}}<b_{Q}+b_{-Q}^{\dagger}>,
\end{eqnarray}
is well-developed in this temperature range. Under such conditions, $b_{1}(q)$ and $b_{2}(q)$ can be identified with the annihilation operators of the amplitude and phase modes of CDW. 
Electron Green's function in the mean field approximation is
\begin{eqnarray}
G(p,i\epsilon_n)=\frac{1}{(i\epsilon_n-\frac{\xi_{p}^2}{4\epsilon_{\rm{F}}})^2-\xi_{p}^2-\Delta^2}\nonumber\\
\times
\left(
\begin{array}{cc}
i\epsilon_n- \frac{\xi_{p}^2}{4\epsilon_{\rm{F}}}-\xi_{p} & \Delta \\
\Delta & i\epsilon_n- \frac{\xi_{p}^2}{4\epsilon_{\rm{F}}}-\xi_{p} 
\end{array}
\right)
\end{eqnarray}
 Collective modes are described by the following Green's functions
\begin{eqnarray}
D_{\alpha}(q,i\omega_n)=-\int_{0}^{\beta}d\tau e^{{\rm i}\omega_n\tau}\Large{<}T_{\tau}\alpha_{q}(\tau)\alpha_{-q}(0)\Large{>},\\
D_{\phi}(q,i\omega_n)=-\int_{0}^{\beta}d\tau e^{{\rm i}\omega_n\tau}\Large{<}T_{\tau}\phi_{q}(\tau)\phi_{-q}(0)\Large{>},
\end{eqnarray}
here, $\alpha_{q}(\tau)$ and $\phi_{q}$ are displacement operators of amplitude and phase modes given by
\begin{eqnarray}
\alpha_{q}=b_{1}(q)+b_{1}^{\dagger}(-q),\\
\phi_{q}=b_{2}(q)+b_{2}^{\dagger}(-q).
\end{eqnarray}
In the low energy region, above Green's function can be written  in \cite{Lee}
\begin{eqnarray}
D_{\alpha}(q,i\omega_n)=\frac{2\Omega}{(1+\frac{1}{3\mu})\omega^2-\lambda\Omega^2-\frac{1}{3\mu}(v_{F}q)^2}, \\
D_{\phi}(q,i\omega_n)=\frac{2\Omega}{(1+\frac{1}{\mu})\omega^2-\frac{1}{\mu}(v_{F}q)^2},
\end{eqnarray}
where $\Omega$ is $\omega_{\pm 2k_{F}}$, $\lambda$ is dimensionless coupling constant $\frac{\gamma^2 a}{\Omega \pi v_F}$, $a$ is lattice constant, and $\mu=\frac{4\Delta^2}{\lambda\Omega^2}$ is the CDW mass paramter. Ordinarily it has quite large values typically on the order of $10^2$. 
\section{Thermoelectric power }
\begin{figure}[htbp]
\includegraphics[width=0.7\linewidth,height=3cm]{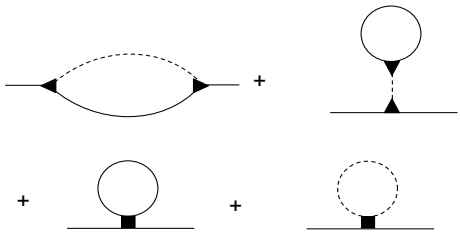}
\caption{Feynman diagrams which give phason self-energy. Each solid and dashed lines show contribution of phase mode and amplitude mode. Electron vertex is shown as triangle and square points whose contributions result in constant values.  }
\label{picture}
\end{figure}

Now let us calculate thermoelectric power.
In this study, we have a interest to collective mode response, so we don't consider impurity effects. We also neglect effect of quasi-particles. Needless to say, these have much effect in real systems, but at least, these contributions are expected to be vanished at low temperatures $T<<T_{\rm{P}}$. Damping is derived from nonlinear interaction between phase mode and amplitude mode, which had been developed by one of the author \cite{Kurihara1, Kurihara2}, yielding the following expression for the phason life time from Fig. 1  
\begin{eqnarray}
\tau=\frac{16\sqrt{\mu}}{\pi\omega_{\alpha}^2}T{\rm sh^2}(\frac{\omega_{\alpha}}{2T}), \label{lifetime}
\end{eqnarray}
where, $\omega_{\alpha}=\sqrt{\frac{3\lambda}{2}}\Omega$. Eq. (\ref{lifetime}) shows that $\tau$ is inversely proportional to $T$ at temperatures above amplitude gap $\omega_{\alpha}$, and diverge exponentially in the zero temperature limit.
\vspace{0.6cm}
\begin{figure}[htbp]
\includegraphics[width=0.6\linewidth,height=3.5cm]{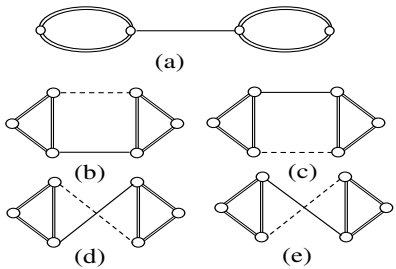}
\caption{Feynman diagrams which give transports coefficients. Double lines represent electrons. One-loop-type diagram $(a)$ contributes to electric conductivity and thermoelectric power, while Aslamazov-Larkin diagrams $(b)\sim (e)$ contribute to electron part of thermal conductivity.    }
\label{picture2}
\end{figure}

In the previous paper  \cite{Kurihara1}, electrical conductivity $\sigma$ is calculated from the diagram shown in Fig.2 (a), and is given by $\sigma=\frac{\omega_{P}^2}{4\pi\mu}\tau$.
This diagram also contributes to other transport coefficients $L^{12}$ and $L^{22}$.  We denote these contribution as $L^{12}_1$ and $L^{22}_1$ which are given by
\begin{eqnarray}
L^{12}_1=\frac{\Delta}{2e}(1-\frac{1}{\lambda})\frac{\Delta}{\epsilon_{F}}L^{11} \label{L121} \label{L12}\\
L^{22}_1=\Big(\frac{\Delta}{2e}(1-\frac{1}{\lambda})\frac{\Delta}{\epsilon_{F}}\Big)^2L^{11} \label{L221}
\end{eqnarray}
There is a factor $\frac{\Delta}{\epsilon_F}$ in above two equations. This means that $L^{12}$ and $L^{22}$ can be non-vanishing only when the electron-hole symmetry is violated.
It is derived from asymmetry between electric and heat current operators. Note that, the factor $\frac{\Delta}{\epsilon_F}$ is not necessarily small in CDW system : a typical value is on the order of 0.1.

 Using Eq. (4), (5) and (\ref{L12}), see that the thermoelectric power is given by
\begin{eqnarray}
S = \frac{\Delta}{2e}(1-\frac{1}{\lambda})\frac{\Delta}{\epsilon_F}\frac{1}{T}\label{tep}
\end{eqnarray}
In the experiment on  $\rm{TaS_{3}}$, field dependence of thermoelectric power 
corresponded to differential resistivity \cite{Stokes}.
This suggest that $L^{12}$ doesn't have dependence on collective modes.
However, in the system of the blue bronze, nonlinear field dependence of $L^{12}$ has observed at several temperatures \cite{Kriza}.
In this system, temperature dependence of thermoelectric power which is shown in Eq. (23) is expected to be observed under the high electric field: above the threshold. We also note that Eq. (\ref{tep}) diverges at zero temperature, which may be an artifact of our assumption: we have ignored impurity effects and quasi particle contributions.
\section{Thermal conductivity }
Thermal conductivity $\kappa$ is composed of two parts. We denote the contribution of electron as $\kappa_{\rm{el}}$ and that of phonon as $\kappa_{\rm{ph}}$

First, we investigate the electron part.
We see that diagram shown in Fig.2 (a) doesn't contribute to collective mode response of thermal conductivity. Actually we can easily confirm the relation $L_{1}^{22}-(L_{1}^{12})^2/L^{11}=0$ from Eq. (\ref{L121}) and Eq. (\ref{L221}). This represents that phase mode transport which is damped by nonlinear interaction between phase mode and amplitude mode is completely canceled out by back flow, i.e. contribution of thermoelectric power and electric current, which is shown in Eq. (6). 

Thus, we investigate other processes of the thermal conductivity. As for higher order diagrams, we focus on Aslamazov-Larkin-type diagrams. Note that Maki-Thompson and density of state diagrams are both important factor in considering the superconducting fluctuations. But these terms give no contribution in our case. We find that Fig.2 $(b)\sim (e)$ give finite values to thermal conductivity, and the other diagrams vanish. As for electrical conductivity and thermoelectric power, all diagrams in this order don't have finite value.

 Actual calculation of these diagrams is rather complicated. We calculate these terms by the following way.
\begin{figure}[htbp]
\includegraphics[width=0.5\linewidth,height=3cm]{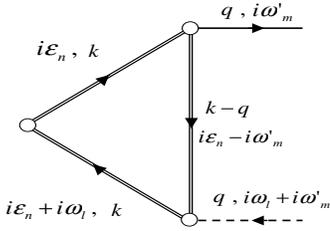}
\caption{An example of diagrams which are shown in FIG. 2. Suffix $k$ and $q$ are momentum, ${\rm i}\epsilon_n$, ${\rm i}\omega_m$ and ${\rm i}\omega_l$ are Matsubara frequency. ${\rm i}\omega_l$ is analytically continuated to $\omega + {\rm i}0$.}
\label{picture2}
\end{figure}
  Summation of these terms are represented as
 \begin{eqnarray}
 & &\frac{1}{\beta N}\sum_{q,{\rm i}\omega_{n}}f(v_{F}q,{\rm i}\omega_{n})[D_{\phi}(v_{F}q,{\rm i}\omega_{n})D_{\alpha}(v_{F}q,{\rm i}\omega_{n}+{\rm i}\omega_{l}) \nonumber \\
 & &+D_{\alpha}(v_{F}q,{\rm i}\omega_{n}+{\rm i}\omega_{l})D_{\phi}(v_{F}q,{\rm i}\omega_{n})]   \label{ph-ap}, 
\end{eqnarray}
where $N$ is number of lattice site, and $\beta=1/T$, function $f(q,{\rm i}\omega_{n})$ is contribution of electron parts shown in Fig. 3.
 As is shown in Fig. 3, we have to consider two summations over Matsubara frequencies ${\rm i}\epsilon_n$ and ${\rm i}\omega_m$. However, we only pick up poles of Green's functions of phonon with respect to ${\rm i}\omega_m$, since we consider the temperature sufficiently below the CDW gap, and effect of quasi particle excitation is expected to be very small. In this assumption, transport coefficient derived from Eq. (\ref{ph-ap}) is given by
\begin{eqnarray}
-\mu^{\frac{1}{2}}\frac{\omega_{\alpha}}{\gamma^2}\frac{8}{T\sinh(\frac{\omega_{\alpha}}{2T})^2}{\large[}f(\sqrt{6}\Delta,\omega_{\alpha})+f(\sqrt{6}\Delta,-\omega_{\alpha}){\large]} \label{ph-ap2}
\end{eqnarray}
Function $f(\sqrt{6}\Delta,\omega_{\alpha})$ is still hard to calculate directly. However, we find $f(0,\omega_{\alpha})+f(0,-\omega_{\alpha})\simeq -(\frac{\omega_{\alpha}}{4\epsilon_F} )^2\frac{\lambda\omega_{\alpha}^2}{3}(1+o(\frac{\omega_{\alpha}}{\Delta}))$. Thus we use this value and finally find to be
\begin{eqnarray}
\kappa_{\rm{el}}\simeq\Big(\frac{3\Delta}{2\epsilon_F}\Big)^2(\frac{\omega_P}{e})^2\frac{3}{4\lambda\pi T}\frac{\hbar}{\tau} \label{kel}.
\end{eqnarray}
 
 Next, we calculate phonon part.
 We find Eq. (\ref{phonon}) represents decay process between phase and amplitude mode, and it's correlation function is composed of product of Green's function of phase mode and amplitude mode. We calculate one-loop contribution of thermal conductivity of phonon part which is given by 
 \begin{eqnarray}
 \kappa_{\rm{ph}}=\Big(\frac{v_{ph}}{v_F}\Big)^2\Big(\frac{\omega_P}{e}\Big)^2\frac{4\mu^{\frac{1}{2}}}{\lambda\pi T}\frac{\hbar}{\tau} \label{kph}.
 \end{eqnarray}
 
$\kappa_{\rm{el}}$ is similar to $\kappa_{\rm{ph}}$, both of them are inversely proportional to phason life time $\tau$.
   Actually, we substitute $f(v_{F}q,{\rm i}\omega_{n}/2)$ with $({\rm i}\omega_n+{\rm i}\omega_l)^2$, then $\kappa_{\rm{ph}}$ can be derived.
We may consider that these heat currents are essentially same: they are carried by energy-interchange process between phase and amplitude modes. Such interpretation is reasonable, because the heat transfer is process of entropy diffusion.  If local amplitude mode excitation causes phase mode excitation and, thereafter phase mode bring about amplitude fluctuation in some other place, it is interpreted as heat current.
  
Finally we estimate the relative importance of these contributions to thermal conductivity.
 $\kappa_{\rm{el}}$ and $\kappa_{\rm{ph}}$ has exponentially low value at temperatures sufficiently below amplitude gap while they have nearly constant value $\acute{\kappa}_{{\rm{el}}}$ and $\acute{\kappa}_{\rm{ph}}$ at temperatures above it.
Ratio between these values and electrical conductivity for this temperature range is given by
\begin{eqnarray}
\acute{\kappa_{\rm{el}}}/\sigma\sim\frac{1}{2\lambda}\Big(\frac{3\Delta}{2\epsilon_{\rm{F}}}\Big)^2L_{0}T \label{kels},\\
 \acute{\kappa_{\rm{ph}}}/\sigma=\frac{2\mu}{3\lambda}\Big(\frac{v_{\rm{ph}}}{v_{\rm{F}}}\Big)^2L_{0}T \label{kphs},
\end{eqnarray}
where $L_{0}$ is Lorentz number $\frac{\pi^2}{3}(k_{\rm{B}}/e)^2$, $k_{\rm{B}}$ is Boltzmann constant.
Since a factor $(\Delta/\epsilon_{\rm{F}})^2$ is quite small which has an order of $10^{-2}$, electonic contribution is small. On the other hand, phonon contribution is much larger, because $\frac{\mu}{\lambda}(v_{\rm{ph}}/v_{\rm{\rm{F}}})^2$ is expected to have an order of 0.1, where we estimate  $v_{\rm{ph}}/v_F$ have an order of $10^{-2}$. 
 \section{Conclusion}
 In conclusion, we have studied thermoelectric properties of a charge density wave system. We have particularly focused on collective mode response.
 We have found temperature dependence of correlation function between electrical current and heat current is same as electrical conductivity, which means thermoelectric power is inversely proportional to temperature. We have also found that thermal conductivity is calculated by energy-interchange process between phase mode and amplitude mode, and is inversely proportional to phason life time.
 \section{acknowledgment}
 We would like to thank Professor I. Terasaki and Professor H. Matsukawa for usefull comments. I also would like to thank Y. Watanabe for stimulating discussions. The work is partly supported by a Grant for The 21st Century COE Program (Holistic Research and Education center for Physics of Self-organization Systems) at Waseda University from the Ministry of Education, Sports, Culture, Science and Technology of Japan.

\newpage
\end{document}